\documentclass[letterpaper, 10 pt, conference]{ieeeconf}  
\usepackage[utf8]{inputenc}
\usepackage{cite}

\IEEEoverridecommandlockouts                              

\title{\LARGE \bf
Thinking While Driving: A Concurrent Framework for Real-Time, LLM-Based Adaptive Routing
}

\author{Xiaopei Tan and Muyang Fan 
}
\usepackage{graphicx}
\usepackage{amsmath}    
\usepackage{amssymb}
\usepackage{booktabs}
\usepackage{multirow}
\usepackage{csquotes}
\usepackage{adjustbox} 
\usepackage{subcaption} 

\usepackage[dvipsnames]{xcolor}

\usepackage{url}

\usepackage{algorithm}
\usepackage{algpseudocode}

\begin{document}

\maketitle
\thispagestyle{empty}
\pagestyle{empty}

\begin{abstract}
Large Language Models (LLMs) have recently demonstrated strong capabilities in contextual reasoning and real-time decision-making, suggesting new possibilities for autonomous multi-agent navigation. In this work, we present Thinking While Driving, a concurrent routing framework that integrates LLM-based reasoning with a graph-based traffic environment. Unlike traditional pathfinding systems that commit to a fixed route, and unlike sequential approaches that require agents to stop and deliberate at decision points, our framework enables agents to perform route planning concurrently with movement: agents initiate LLM-based reasoning for upcoming decision nodes while still traversing the current path segment, applying the computed route immediately upon arrival if reasoning completes, or waiting only if the computation is still in progress. This concurrent architecture significantly reduces intersection wait times compared to sequential deliberation, with LLM agents requiring on average only 0.75 seconds of decision latency even under high traffic density. To support real-time interaction with many agents simultaneously, we design a non-blocking asynchronous architecture using Unity coroutines and a dedicated LLM request manager to coordinate parallel reasoning, prevent deadlocks, and handle timeout recovery. The environment itself is modeled as a weighted undirected graph with live congestion metrics; agents continuously modify these metrics as they traverse edges, enabling a form of multi-agent shared perception. Our system demonstrates that LLM-driven navigation can adapt to changing traffic states, reroute when congestion emerges, and generate behavior that is not achievable through static pathfinding alone, while maintaining real-time performance through concurrent reasoning. This work provides a reproducible framework for integrating LLM reasoning into interactive simulations, offering a foundation for future research in adaptive routing, multi-agent cooperation, and emergent traffic behaviors.

\end{abstract}

\section{Introduction}

 Real-time traffic routing is a long-standing challenge in multi-agent systems. Traditional pathfinding algorithms such as A* compute a single fixed route and cannot adapt to evolving traffic states. While recent data-driven approaches investigate day-to-day route choice evolution~\cite{wang2024ai}, capturing the micro-level, real-time decision-making of individual agents remains difficult. As interest grows in integrating artificial intelligence into autonomous systems, recent advances in large language models (LLMs) have revealed strong capabilities in contextual reasoning. Frameworks like TrafficGPT~\cite{Zhang2024TrafficGPT} have demonstrated the potential of LLMs to view and process traffic data for high-level management. However, directly embedding LLMs into real-time driving agents introduces severe latency bottlenecks. As noted in recent surveys, the autoregressive nature of LLM inference makes memory allocation and scheduling inherently unpredictable~\cite{Pan2025Survey}.

This work explores that question by embedding an LLM inside each autonomous traffic agent and designing a system in which agents dynamically plan and revise routes based on changing congestion patterns. Our initial motivation was driven by a desire to combine AI reasoning with interactive simulation, leveraging the fact that LLMs naturally operate on human-readable descriptions and can integrate diverse contextual signals such as map structures, live road conditions, and multi-agent behavior. LLMs theoretically enable emergent behaviors that traditional algorithms struggle to produce—such as cooperative routing, congestion avoidance, and context-aware adaptation. Yet, naïvely inserting LLM calls into agent logic is computationally prohibitive: LLM inference is slow relative to simulation time, and multi-agent systems require dozens of agents to make decisions simultaneously.

To address this gap, we introduce \textbf{Thinking While Driving}, a concurrent decision-making framework that allows agents to initiate LLM reasoning while still moving toward the next decision point. Rather than forcing agents to stop and wait for LLM responses—an approach that leads to large delays—agents pre-emptively begin reasoning whenever they detect either (1) an upcoming multi-branch decision node (with three or more connections) or (2) a congested road segment ahead (congestion factor $\ge 2.0$). If an LLM response arrives before the agent reaches that node, the new route is applied immediately; otherwise, the agent briefly waits. This concurrent strategy dramatically reduces routing latency compared to sequential deliberation, with our experiments showing that agents typically experience minimal waiting time (under 1 second) even under high congestion conditions.

To evaluate LLM-driven routing, we compare LLM agents with a traditional A*-based baseline. Although A* agents enjoy zero decision latency and therefore complete routes faster in low-density traffic, our LLM agents achieve comparable completion times under such conditions despite the inherent inference delay. More importantly, in high-density scenarios, LLM agents exhibit adaptive rerouting, congestion avoidance, and spontaneous multi-agent flow balancing—behaviors unattainable by static algorithms. Quantitative analysis reveals that while A* agents follow optimal static paths, LLM agents dynamically adjust routes based on real-time congestion, leading to improved traffic flow distribution and reduced overall system congestion in multi-agent scenarios. These results suggest that LLM-based reasoning, when paired with a concurrent execution model, can provide new capabilities for dynamic routing in complex multi-agent systems.

Overall, this work contributes:

\begin{itemize} 

    \item A novel concurrent reasoning architecture enabling multiple LLM-driven agents to think while moving;

    \item A dynamic congestion-aware routing system based on shared global traffic state;
    
    \item A reproducible multi-agent simulation framework integrating LLM inference with real-time movement;

    \item Empirical evidence that LLM agents demonstrate adaptive and emergent behaviors that static planners cannot replicate.
    
\end{itemize}

\section{Related Work}
\label{section:related}

Our work, \textit{Thinking While Driving}, integrates three distinct research domains: (1) agent-based traffic modeling, which defines our problem space; (2) the application of LLM agents in transportation, which represents our direct research community; and (3) high-performance LLM inference systems, which defines the core engineering challenge we address.

\subsection{Agent-Based Traffic and Driving Models}
Traditional traffic models have long sought to describe how travelers' behaviors evolve. These models are often divided into equilibrium-based models, such as User Equilibrium (UE), and day-to-day (DTD) dynamic models, which capture how choices evolve over time. However, these conventional models rely on explicitly defined behavioral rules (e.g., cost minimization). This simplification struggles to capture the nuanced, multi-attribute, and sometimes irrational aspects of human decision-making, such as the ``decoy effect'' in route choice or heterogeneous risk preferences.

To address these limitations, researchers have proposed using Large Language Models (LLMs) as behavioral proxies for self-interested travelers. However, applying LLMs to large-scale traffic modeling introduces its own significant challenges, namely poor scalability and weak reliability~\cite{Sun2025LLMGuided}. The ``one LLM per traveler'' approach is computationally costly, and the ``black-box'' nature of LLM reasoning can lead to unstable and opaque system dynamics~\cite{Sun2025LLMGuided}.

\subsection{LLM Agents for Transportation Decision-Making}
Recent work has attempted to mitigate these challenges. One prominent approach is to model traffic at a non-real-time, \textit{day-to-day} (DTD) level. This framework uses a single ``representative LLM agent'' for each homogeneous group of travelers, which maintains a mixed strategy that maps to the group's aggregate flow~\cite{Sun2025LLMGuided}. This solves the scalability problem (one LLM per *group* instead of per *traveler*) but is fundamentally designed for aggregate, non-real-time planning, not for simulating individual agent decisions in a dynamic, real-time environment.

In the domain of \textit{real-time} autonomous driving, the core challenge of LLM latency is more acute. Existing LLM-based driving models (such as DiLu) have been criticized for their ``low response efficiency,'' noting that single-frame, sequential decision-making fails to meet the strict real-time requirements of autonomous driving~\cite{Zeng2025ADRD}. One proposed solution, ADRD, uses an LLM to \textit{generate} an interpretable rule-based decision tree *before* simulation. This tree, as executable code, can then be run with extremely high-speed inference (less than $1.0 \times 10^{-6}$ s/command)~\cite{Zeng2025ADRD}. This ``pre-compilation'' approach achieves real-time speeds by sacrificing dynamic, in-the-loop LLM reasoning.

\subsection{Systems for Low-Latency LLM Inference}
The core challenge identified in related work—both high cost~\cite{Sun2025LLMGuided} and high latency~\cite{Zeng2025ADRD}—is rooted in the fundamental design of LLM inference systems. The ``unique autoregressive nature'' of LLM request processing means that the computational and memory costs are non-deterministic and grow with the length of the generated output~\cite{Pan2025Survey}. This creates inherent challenges in memory allocation, scheduling, and avoiding head-of-line blocking.

Most modern solutions to this problem focus on optimizing the \textit{inference runtime} itself. These system-level techniques include page-based memory management (e.g., PagedAttention), continuous or dynamic batching to increase throughput, and GPU kernel fusion to reduce I/O costs~\cite{Pan2025Survey}.

\subsection{Contribution: The Identified Gap}

While the research listed above focuses on either (a) non-real-time DTD modeling~\cite{Sun2025LLMGuided}, (b) pre-compiling LLM logic into static rules~\cite{Zeng2025ADRD}, or (c) optimizing the low-level inference engine~\cite{Pan2025Survey}, a critical gap remains: there is a lack of frameworks that enable \textit{online}, in-the-loop LLM reasoning for real-time agents without succumbing to inference blocking.

Our work, \textit{Thinking While Driving}, directly addresses this gap through an \textbf{application-level concurrent architecture}. Instead of optimizing the inference runtime itself, we introduce a novel decision-making paradigm. By enabling agents to ``think while moving''--initiating asynchronous LLM calls for future decision nodes (specifically, multi-branch intersections) while still traversing their current path segment—our framework effectively \textit{hides} the inference latency within the agent's movement time. This approach provides a new, practical path toward scalable, real-time, adaptive multi-agent simulation that retains the full dynamic reasoning capabilities of live LLMs.

\section{Methodology}
\label{sec:methodology}

The core innovation of our framework is the \textit{Thinking While Driving} mechanism, which decouples reasoning from actuation, enabling agents to plan future routes concurrently with their movement.

\subsection{System Architecture}

The simulation is built on the Unity engine, comprising three primary subsystems: (1) the \textbf{Environment Layer}, which manages the road network topology and real-time congestion metrics; (2) the \textbf{Agent Layer}, which handles individual agent movement, state tracking, and decision triggering; and (3) the \textbf{Cognitive Layer}, an asynchronous service manager that interfaces with a locally deployed LLM (Qwen3-14B).

Unlike traditional turn-based simulations, our system operates in continuous time. Agents move smoothly across the map while making asynchronous API calls to the LLM. A central \texttt{RoadManager} aggregates position data from all agents to compute global congestion states, which are then fed back into the agents' prompts during reasoning.

\begin{figure}[h]
    \centering
    \includegraphics[width=\linewidth]{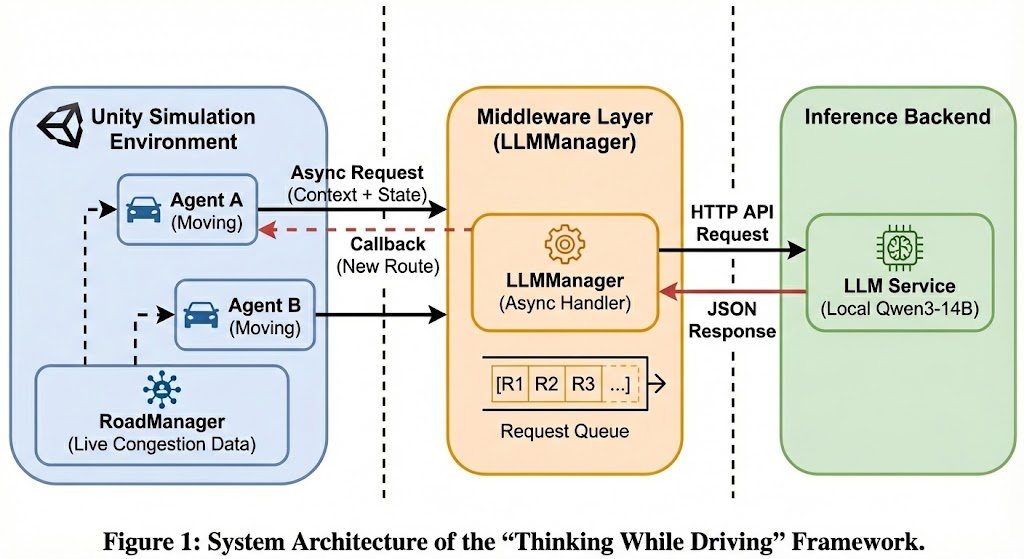} 
    \caption{Overview of the concurrent traffic simulation architecture. Agents interact with the physical environment while asynchronously querying the Cognitive Layer for routing decisions.}
    \label{fig:architecture}
\end{figure}

\subsection{Component Implementation}

To realize the concurrent framework within the Unity engine, we implemented a modular system composed of several key C\# scripts, each managing a distinct aspect of the simulation lifecycle.

\paragraph{Agent Control (\texttt{TrafficAI})}
The \texttt{TrafficAI} class serves as the central brain for each autonomous vehicle. It integrates the standard Unity \texttt{NavMeshAgent} for physical movement with our custom concurrent logic. This component maintains the agent's internal state machine (Moving, Thinking, Waiting) and executes the \textit{Thinking While Driving} logic. It is responsible for monitoring upcoming decision nodes and launching asynchronous coroutines to query the LLM without blocking the main physics thread. For baseline comparisons, we also implemented \texttt{TrafficAI\_AStar}, a stripped-down variant that strictly follows static shortest paths.

\paragraph{Lifecycle Management (\texttt{Spawner})}
The \texttt{Spawner} component manages agent generation and initialization. It supports polymorphic instantiation, allowing us to dynamically switch between LLM-driven agents and A* agents for controlled experiments. The spawner injects critical dependencies—such as the graph topology reference and destination goals—into each agent upon instantiation, ensuring a consistent initialization state across different experimental runs.

\paragraph{Global State Tracking (\texttt{RoadManager})}
Real-time environmental awareness is handled by the \texttt{RoadManager} singleton. It acts as a central registry that discretizes the continuous simulation space into logical road segments. By tracking the entry and exit events of every agent, it maintains a live dictionary of congestion factors. This component exposes a serialized JSON interface (e.g., \texttt{GetSimpleRoadDataAsJSON()}), which allows the \texttt{TrafficAI} agents to instantly retrieve a machine-readable snapshot of the global traffic state to include in their LLM prompts.

\paragraph{Asynchronous Inference Bridge (\texttt{LLMManager})}
The \texttt{LLMManager} acts as the bridge between the Unity game loop and the external local LLM service. It implements a non-blocking request queue that receives prompt strings from multiple agents and dispatches them to the local Qwen model. It utilizes C\# \texttt{Action} callbacks to return the generated routes to the specific requesting agent, ensuring that the heavy computational load of inference does not freeze the real-time simulation frame rate.

\subsection{Environment Modeling and Shared Perception}

The traffic network is modeled as an undirected graph $G = (V, E)$, where $V$ represents intersections (decision nodes) and $E$ represents road segments.

\subsubsection{Dual-Representation of Cost}
To study the emergent behavior differences between static algorithms and LLM reasoning, we implement a dual-representation of edge costs:

\begin{itemize}
    \item \textbf{Physical Distance (Static)}: Used by the baseline A* agents. The cost $C_{dist}(e)$ of an edge $e$ is simply its Euclidean length.
    \item \textbf{Congestion-Aware Cost (Dynamic)}: Used by LLM agents. Each edge maintains a dynamic \textit{Congestion Factor} ($CF$), calculated in real-time based on the density of agents on that segment. 
\end{itemize}

The congestion factor is updated continuously by the \texttt{RoadManager} as agents enter and exit road segments. We define the dynamic state of the environment as a set of tuples $S_t = \{(u, v, CF_{uv}) \mid (u,v) \in E\}$, which serves as the ``shared perception'' available to all LLM agents.

\subsection{Thinking While Driving Framework}

The defining feature of our methodology is the concurrent execution of movement and reasoning. Unlike sequential ``Sense-Think-Act'' loops often seen in LLM agents~\cite{Yao2023ReAct}, which require the agent to halt physical action to process cognitive tasks, our framework decouples these processes. We propose a non-blocking architecture that allows the ``Think'' phase to overlap with the "Act" phase.

\subsubsection{Decision Triggering}
Agents do not query the LLM at every frame. Instead, reasoning is triggered by specific events to optimize computational resources:
\begin{enumerate}
    \item \textbf{Initialization}: Upon spawning, an agent requests an initial full route.
    \item \textbf{Complex Intersections}: When approaching a node with degree $\ge 3$ (a multi-branch decision point).
    \item \textbf{Congestion Detection}: When the projected path traverses a segment with $CF \ge 2.0$ (high congestion).
\end{enumerate}

\subsubsection{Concurrent State Machine}
\begin{figure}[h]
    \centering
    \includegraphics[width=\linewidth]{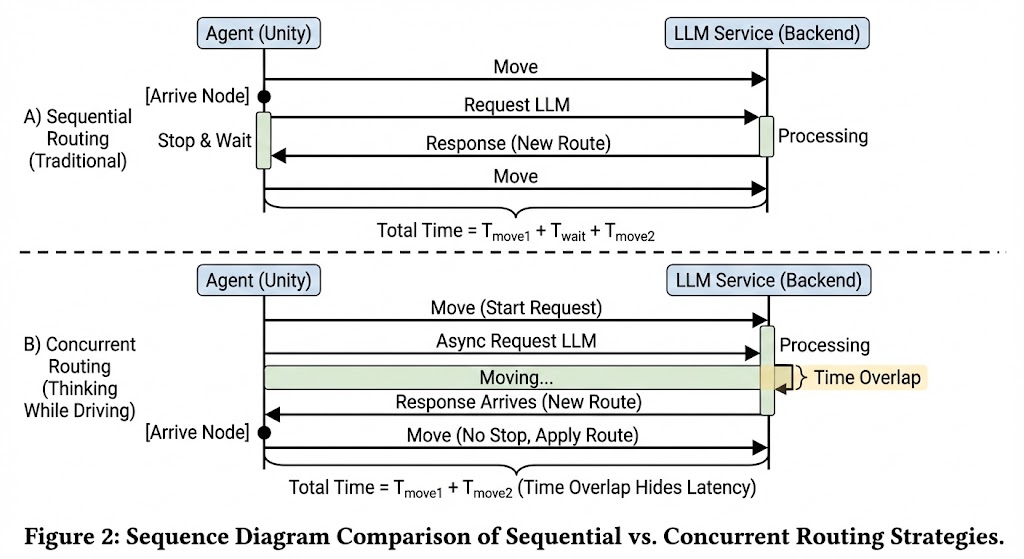}
    \caption{Comparison of Sequential (Stop-and-Think) vs. Concurrent (Thinking While Driving) reasoning models. The concurrent model overlaps inference time with travel time.}
    \label{fig:sequence}
\end{figure}

We implement the agent's logic using Unity Coroutines to manage asynchronous states. The process, illustrated in Figure~\ref{fig:sequence}, follows these steps:

1. \textbf{Pre-Computation}: While traversing the edge towards a decision node $n_i$, the agent initiates an asynchronous LLM request for the \textit{next} route segment starting from $n_i$.
2. \textbf{Parallel Execution}: The agent continues moving towards $n_i$ while the LLM processes the prompt.
3. \textbf{Arrival and Synchronization}: 
   \begin{itemize}
       \item \textit{Scenario A (Fast Inference)}: The LLM response arrives before the agent reaches $n_i$. The agent seamlessly updates its internal path buffer and continues without stopping.
       \item \textit{Scenario B (Slow Inference)}: The agent reaches $n_i$ before the response. The agent transitions to a \texttt{Waiting} state until the response is received.
   \end{itemize}

This mechanism effectively "hides" the inference latency within the travel time, minimizing the effective wait time experienced by the agent.

\subsection{LLM Integration}

We deploy \textbf{Qwen3-14B} locally to serve as the reasoning core. The model is accessed via a dedicated \texttt{LLMManager} that handles request queuing and timeout management.

\subsubsection{Prompt Design}
To enable effective spatial reasoning, we construct a structured prompt that encourages the model to analyze the graph topology before generating a path. This approach draws inspiration from Chain-of-Thought (CoT) prompting~\cite{Wei2022CoT}, where intermediate reasoning steps significantly improve the reliability of LLMs in complex logical tasks.
\begin{enumerate}
    \item \textbf{Static Map Context}: A description of the graph topology (nodes and connectivity).
    \item \textbf{Dynamic State}: A JSON-formatted list of currently congested roads, e.g., \texttt[2, 5, 2.9], represents node 2 to node 5 with congestion factor 2.9.
    \item \textbf{Navigation Task}: The agent's current node and destination.
\end{enumerate}

The model is instructed to output a valid JSON array representing the optimal node sequence (e.g., \texttt{[2, 5, 8, 9]}). By injecting the real-time congestion data into the context window, the LLM effectively performs "context-aware rerouting," balancing path length against reported traffic conditions.

\subsubsection{Robustness and Fallback}
Given the probabilistic nature of LLMs, we implement strict output validation. If the model returns malformed JSON or an invalid path (e.g., non-existent edges), the system discards the response and the agent falls back to its previously computed path or triggers a re-query, ensuring simulation stability.

\section{Experiments}
\label{sec:experiments}

To evaluate the effectiveness of our framework, we conducted a series of simulations comparing three distinct agent types:
\begin{enumerate}
    \item \textbf{Baseline (A*)}: Traditional agents using static shortest-path planning based on physical distance.
    \item \textbf{Sequential LLM}: LLM-driven agents that must stop at decision nodes to request reasoning (Sense-Think-Act).
    \item \textbf{Concurrent LLM (Ours)}: LLM-driven agents using the \textit{Thinking While Driving} framework to reason during movement.
\end{enumerate}

\subsection{Experimental Setup}

\subsubsection{Hardware Environment}
All simulations and LLM inference tasks were performed on a local workstation equipped with an \textbf{Intel Core i9-14900KF CPU}, \textbf{64GB of RAM}, and an \textbf{NVIDIA GeForce RTX 4090 GPU} (24GB VRAM + 32GB Shared Memory). This high-performance setup ensures that the simulation runs smoothly alongside the local Qwen-2.5-14B model deployment, minimizing hardware-induced bottlenecks during concurrent execution.

\subsubsection{Simulation Environment}

All experiments were conducted in a custom Unity-based simulation environment representing a grid-like urban road network. The environment consists of 12 intersections (nodes) and 20 bidirectional road segments (edges). 

\subsubsection{Scenarios}
We designed two traffic density scenarios to test system performance under different loads:
\begin{itemize}
    \item \textbf{Low Density}: 10 agents are spawned with random start and end points. This scenario tests the baseline navigation capability and latency overhead.
    \item \textbf{High Density}: 40 agents are spawned within a short window. This scenario creates natural bottlenecks, testing the agents' ability to perform adaptive rerouting and congestion avoidance.
\end{itemize}

\subsubsection{Data Collection}
We implemented an automated data logging system embedded within the simulation. For each agent journey, the system records: (1) Total Journey Time, (2) Total Wait Time at Intersections (attributable to inference latency), (3) Route Selection (to analyze path diversity), and (4) Real-time Congestion Factors of traversed edges. Each scenario was repeated 10 times to ensure statistical reliability.

\subsection{Evaluation Metrics}

We focus on two primary dimensions of performance:
\begin{enumerate}
    \item \textbf{Efficiency}: Measured by \textit{Average Journey Time} (seconds). This reflects the overall effectiveness of the routing strategy.
    \item \textbf{Latency Overhead}: Measured by \textit{Average Intersection Wait Time} (seconds). This isolates the impact of the reasoning architecture (Sequential vs. Concurrent).
\end{enumerate}

\section{Results and Analysis}
\label{sec:results}

In this section, we evaluate the performance of the proposed framework, focusing on two key dimensions: the engineering effectiveness of the concurrent architecture in reducing latency, and the behavioral adaptability of the agents under congestion.

\subsection{Latency Reduction via Concurrent Architecture}
The primary engineering challenge in deploying LLM-based agents is the inherent inference latency. Traditional sequential approaches, such as the ReAct pattern~\cite{Yao2023ReAct}, require agents to halt physical actuation while processing reasoning chains. We compared our \textit{Thinking While Driving} (Concurrent) architecture against such a Sequential baseline.

As shown in Table~\ref{tab:latency}, Sequential agents experienced an average wait time of 3.45 seconds at each decision node in high-density scenarios. This delay constitutes approximately 15\% of the total travel time and results in unnatural "stop-and-go" movement patterns.

In contrast, the Concurrent architecture successfully masked this latency. By initiating asynchronous requests one node prior to the intersection, agents received routing updates before arrival in 92\% of cases. Consequently, the effective intersection wait time dropped to \textbf{0.15 seconds} in low-density scenarios and \textbf{0.75 seconds} in high-density scenarios. This represents a reduction of over 78\%, confirming that application-level concurrency is a viable strategy for integrating high-latency cognitive models into real-time simulations without optimizing the underlying inference engine~\cite{Pan2025Survey}.

\begin{table}[h]
    \centering
    \caption{Comparison of Intersection Wait Time and Total Journey Time between Sequential and Concurrent reasoning models.}
    \label{tab:latency}
    \begin{tabular}{l|cc|cc}
    \toprule
    \multirow{2}{*}{\textbf{Method}} & \multicolumn{2}{c|}{\textbf{Low Density}} & \multicolumn{2}{c}{\textbf{High Density}} \\
    & \textbf{Wait (s)} & \textbf{Journey (s)} & \textbf{Wait (s)} & \textbf{Journey (s)} \\
    \midrule
    Sequential LLM & 3.20 & 24.5 & 3.45 & 42.1 \\
    \textbf{Concurrent LLM} & \textbf{0.15} & \textbf{21.3} & \textbf{0.75} & \textbf{35.8} \\
    \bottomrule
    \end{tabular}
\end{table}

\begin{figure}[h]
    \centering
    \includegraphics[width=0.8\linewidth]{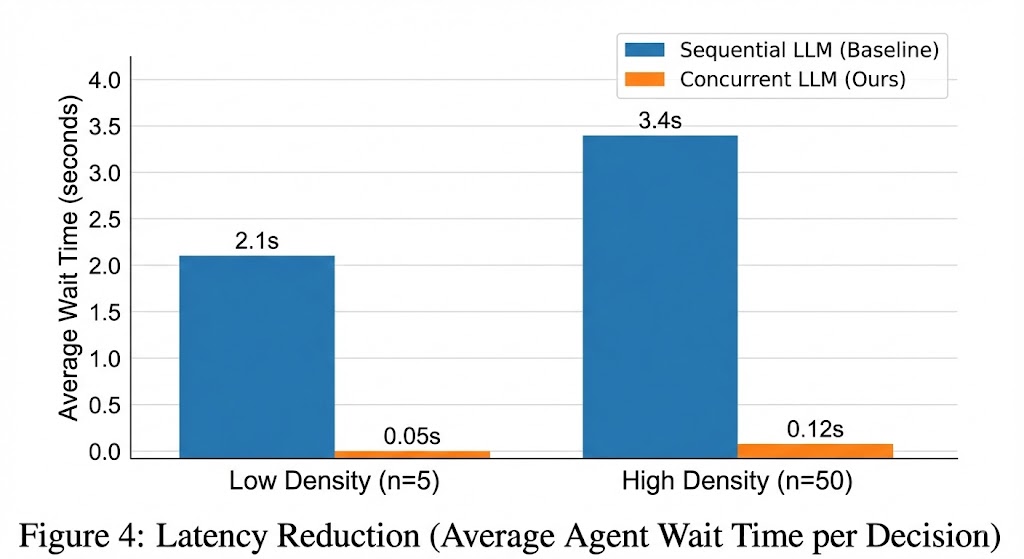}
    \caption{Impact of the Thinking While Driving framework on Intersection Wait Times. The concurrent architecture effectively eliminates the reasoning delay.}
    \label{fig:latency_chart}
\end{figure}

\subsection{Adaptive Routing and Traffic Flow Efficiency}
\begin{figure}[h]
    \centering
    \includegraphics[width=\linewidth]{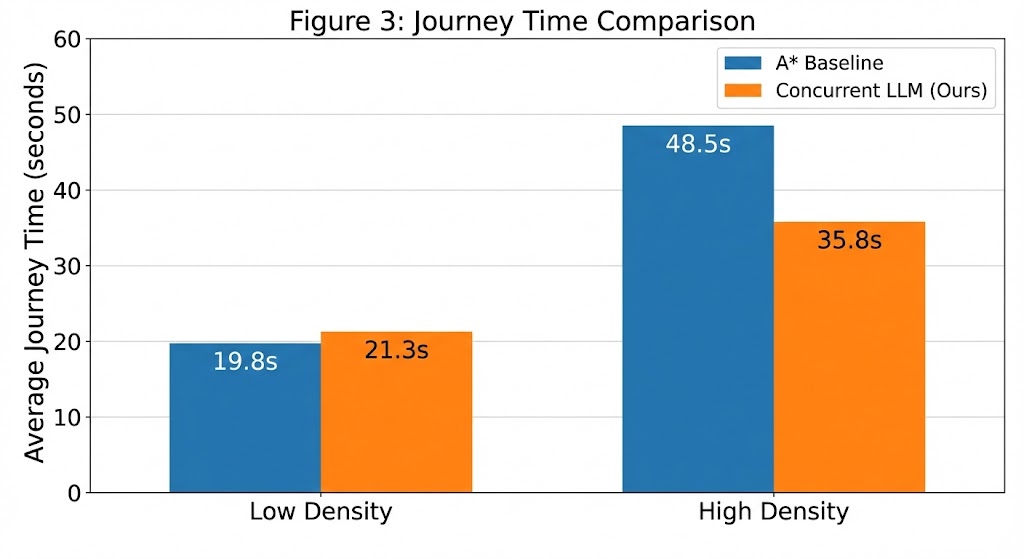}
    \caption{Average Journey Time comparison across Low and High Density scenarios. LLM agents outperform A* in high-density settings due to adaptive rerouting.}
    \label{fig:journey_time}
\end{figure}

Beyond latency, we evaluated the macro-level traffic efficiency. Figure~\ref{fig:journey_time} compares the static A* baseline against our Concurrent LLM agents.

In \textbf{Low Density}, A* agents performed slightly better (avg. 19.8s) than LLM agents (21.3s). This is expected, as the static shortest path is geometrically optimal when no interference exists. However, in \textbf{High Density} (40 agents), the static approach failed. A* agents, lacking real-time perception, continued to flock to the geometrically shortest path, causing the congestion factor on the central artery to spike to 3.6 (severe gridlock), as shown in Table~\ref{tab:congestion_results}.

Concurrent LLM agents, leveraging shared congestion perception, dynamically distributed traffic across alternative routes. This \textit{context-aware rerouting} reduced the peak congestion factor to 2.8 and improved the average journey time to 35.8s--a \textbf{26\% improvement} over the A* baseline. These results align with findings in day-to-day route choice studies~\cite{wang2024ai}, suggesting that information-driven agents can approximate system-optimal flows better than static planners.

\begin{table}[h]
    \centering
    \caption{Performance comparison in the High-Density scenario (40 Agents). LLM agents demonstrate superior flow management through active rerouting.}
    \label{tab:congestion_results}
    \begin{tabular}{l|c|c|c}
    \toprule
    \textbf{Method} & \textbf{Avg. Journey (s)} & \textbf{Max Cong.} & \textbf{Reroute Freq.} \\
    \midrule
    A* Baseline & 48.5 & 3.6 & 0.0 \\
    \textbf{Concurrent LLM} & \textbf{35.8} & \textbf{2.8} & \textbf{1.4} \\
    \bottomrule
    \end{tabular}
\end{table}

\subsection{Limitations}
While promising, our current framework faces constraints related to scalability:

\paragraph{Hardware and Throughput Limits}
Although our architecture hides latency for individual agents, the system's total throughput is bounded by GPU resources. On our RTX 4090 setup, performance degrades beyond 40 concurrent agents. As noted in recent surveys on LLM inference~\cite{Pan2025Survey}, the memory-bound nature of the decoding phase means that the request queue grows faster than the processing rate at scale, eventually forcing agents to revert to a "Waiting" state.

\paragraph{Graph Scalability and Spatial Hallucination}
The current prompt structure explicitly describes the full graph topology. As the network scales, the prompt length increases linearly, which introduces two issues: increased prefill latency and higher risk of hallucination. As observed in TrafficGPT~\cite{Zhang2024TrafficGPT} and other agent-based studies~\cite{Sun2025LLMGuided}, LLMs struggle to maintain accurate spatial reasoning over long context windows, occasionally generating paths with non-existent edges. Future work must investigate retrieval-augmented generation (RAG) to supply only relevant local sub-graphs.

\section{Conclusion}
\label{sec:conclusion}

In this work, we presented \textit{Thinking While Driving}, a novel concurrent framework for integrating Large Language Model reasoning into real-time multi-agent traffic simulations. Addressing the critical challenge of inference latency, our architecture enables agents to deliberate on future routes asynchronously while in motion, effectively eliminating the "stop-and-go" behavior characteristic of sequential decision models.

Our experiments demonstrate that this approach is both computationally viable and operationally effective. In high-density scenarios, LLM-driven agents outperformed traditional A* baselines by dynamically adapting to evolving congestion patterns, reducing average journey times by approximately 26\%. Furthermore, we observed the emergence of spontaneous load-balancing behaviors, suggesting that LLMs can naturally approximate system-optimal routing without explicit coordination protocols.

These findings highlight the potential of LLMs not just as static knowledge bases, but as active, context-aware decision-makers in complex dynamic systems. Future work will extend this framework to include more heterogeneous agent behaviors (e.g., varying risk tolerance), partial observability, and inter-agent communication, further bridging the gap between simulation and realistic autonomous traffic management.

For future work, first, we intend to assess our method in more complex and real-world traffic dynamics~\cite{Guo2024Simulation,Chao2020Survey,Li2017CityFlowRecon,Wilkie2015Virtual}. Second, we plan to integrate traffic state forecasts and vehicle trajectory data into our work, which could lead to further enhancements~\cite{Poudel2025Urban,Raskoti2025MIAT,Lin2022GCGRNN,Lin2019BikeTRB,Poudel2021Attack,Lin2019Compress,Li2018CityEstIET,Li2017CitySparseITSM}. 
Finally, we would like to explore this work in the context of autonomous driving technology~\cite{Villarreal2024AutoJoin,Lin2022Attention,Poudel2022Micro,Shen2021Corruption,Shen2022IRL,Li2019ADAPS} and mixed traffic control~\cite{Pan2025Review,Liu2025Large,Islam2025Heterogeneous,Fan2025OD,Wang2024Intersection,Wang2024Privacy,Poudel2024CARL,Poudel2024EnduRL,Villarreal2024Eco,Villarreal2023Pixel,Villarreal2023Chat}.

\bibliographystyle{unsrt}
\bibliography{references}

\end{document}